\documentclass[11pt,twoside]{article}

\usepackage{asp2014}

\aspSuppressVolSlug
\resetcounters

\begin{document}

\title{TOPCAT and Gaia}

\author{M.~B.~Taylor
  \affil{H.~H.~Wills Physics Laboratory, Tyndall Avenue, 
         University of Bristol, UK;
         \email{m.b.taylor@bristol.ac.uk}}
}

\paperauthor{M.~B.~Taylor}{m.b.taylor@bristol.ac.uk}{0000-0002-4209-1479}{University of Bristol}{School of Physics}{Bristol}{Bristol}{BS8 1TL}{U.K.}


\begin{abstract}
TOPCAT, and its command line counterpart STILTS, are powerful tools
for working with large source catalogues.  ESA's {\it Gaia\/} mission,
most recently with its second data release, is producing source catalogues
of unprecedented quality for more than a billion sources.
This paper presents some examples of how TOPCAT and STILTS
can be used for analysis of {\it Gaia\/} data.
\end{abstract}


\section{Introduction}

TOPCAT \citep{2005ASPC..347...29T} and
STILTS \citep{2006ASPC..351..666T} are
respectively GUI and command-line analysis packages for working
with tabular data in astronomy, and as such offer
many facilities for manipulation of data such as source catalogues.
The recent second data release from ESA's {\it Gaia\/} mission
\citep{2018A&A...616A...1G}
has produced a source catalogue of exceptional quality,
and TOPCAT/STILTS are well-placed to provide analysis capabilities
for exploitation of this data set.
This paper discusses some of the features of the software
most relevant for working with the {\it Gaia\/} DR2 catalogue.
Some have been added specifically with {\it Gaia\/} data in mind,
but in most cases they are general purpose capabilities that are
also suitable for use with other datasets.

\section{Data Access}

The primary access to the {\it Gaia\/} catalogue is via
{\em Virtual Observatory\/} protocols,
provided from the main archive service at ESAC and
a number of other data centers.

The most capable of these access protocols is TAP, the Table Access Protocol,
which allows execution in the archive database of user-supplied
SQL-like queries and retrieval of the resulting tables.
TAP, while allowing extremely powerful queries to be performed,
is a complex protocol stack which presents some challenges
for the client software and user alike.
TOPCAT provides the user with a GUI client for
interacting with TAP services that integrates functions such as
metadata browsing, query validation, table upload and query submission
to make the use of TAP as straightforward as possible for the user,
without obscuring the flexibility it offers
\citep{2017ASPC..512..589T}.
For simpler queries, a Cone Search client is also provided
for retrieving source lists based on sky position alone.
The {\it Gaia\/} catalogue additionally contains non-scalar data for some rows,
exposed using the VO DataLink protocol.
At DR2 this array data is limited to epoch photometry of
a relatively small number of known variable sources,
but much more, including spectrophotometry, will be provided in
future data releases.
TOPCAT's {\em Activation Action\/} toolbox has been overhauled
in recent versions to work with this array data.

Since these services follow the VO standards,
no {\it Gaia}-specific code is required or implemented in TOPCAT
to provide data access.  This means that the same clients can be used
for working with copies of the {\it Gaia\/} catalogue in the main archive
and elsewhere, as well as with other VO-compliant services.
This standardisation has benefits for both the implementer and user
of the software.

The only truly {\it Gaia}-specific code in TOPCAT is a reader for
the GBIN format used internally by the analysis consortium.
This is a specialised capability of no interest to the general
astronomy user, but it has proved valuable for DPAC members
working with the data prior to catalogue publication.

\section{Expression Language}

TOPCAT provides a powerful language for evaluating algebraic
expressions to define new columns, row selections and plot coordinates.
As well as standard arithmetic, trigonometric and astronomical
operations, the library contains a number of astrometric
functions:
\begin{itemize}
\item Propagation of astrometric parameters to earlier/later epoch,
      with or without errors and correlations
\item Conversion of positions and velocities from astrometric parameters
      to Cartesian coordinates in ICRS, galactic or ecliptic coordinates
\item Bayesian estimation of distances and uncertainties from parallax,
      using the expressions from \citet{2016ApJ...833..119A}
\end{itemize}
These are not exactly specific to {\it Gaia},
but they have been added as they are likely to be often needed when
working with {\it Gaia\/} data, and they are specified and documented
in a way that makes them easy to use in that context.
For instance the following expression calulates the $(U,V,W)$ components
of velocity in the Galactic coordinate system (without adjusting for local
standard of rest, and assuming that parallax error is low):
\begin{verbatim}
   icrsToGal(astromUVW(ra, dec, pmra, pmdec,
                       radial_velocity, 1000./parallax, false))
\end{verbatim}
The variable names here are {\tt gaia\_source} catalogue column names,
and the units are as supplied in the catalogue.

\section{Scalability}

\begin{figure}
\includegraphics[width=\textwidth]{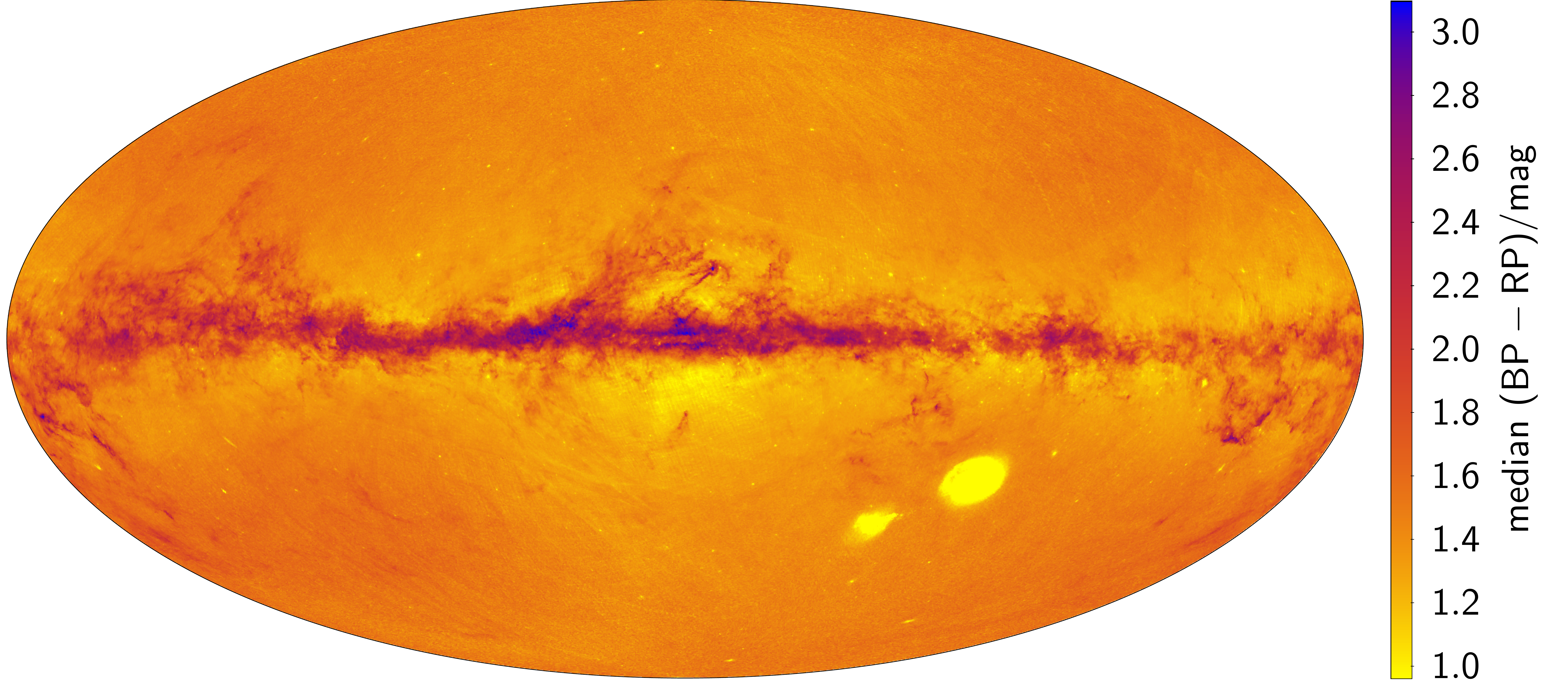}
\caption{All-sky plot of $BP-RP$ colour for 1.4 billion {\it Gaia} sources.
         The colours represent the median within each level 8 HEALPix pixel.
         \label{fig:P3-8:allsky}}
\end{figure}

{\it Gaia\/} DR2 contains 1.7 billion sources, and
investigating this dataset often requires working with large tables.
TOPCAT is well-suited for interactive analysis,
including flexible exploratory visualisation,
of tables (for instance selections from the full catalogue)
with the order of $10^6$--$10^7$ rows.
This regime is quite usable on modest hardware with
no special data preparation,
for instance data downloaded from external services,
or loaded from local FITS or even CSV files.
TOPCAT can be used with tables larger than this,
but interactive performance may be poor or memory exhausted.

STILTS on the other hand processes data for the most part in
streaming mode, so can cope with arbitrarily large tables
in fixed memory.  This means that non-interactive calculations
or preparation of graphics for the entire {\it Gaia\/} catalogue is quite
feasible.
A set of all-sky weighted density maps using all 1.7 billion rows
was prepared as follows:
\begin{enumerate}
\vspace*{-1ex}
\item download 61\,000 (0.5\,Tb) gzipped CSV files from ESAC
      ({\tt wget}, 10 hours)
\vspace*{-1ex}
\item convert to 61\,000 small FITS files
      (STILTS {\tt tpipe}, 5 days)
\vspace*{-1ex}
\item convert to single 0.8\,Tb column-oriented FITS file
      (STILTS {\tt tpipe}, 12 hours)
\vspace*{-1ex}
\item aggregate into level-9 HEALPix map
      (STILTS {\tt tskymap}, 45 minutes)
\vspace*{-1ex}
\item render plot to PDF or PNG
      (STILTS {\tt plot2sky}, a few seconds)
\end{enumerate}
\vspace*{-1ex}
In practice, steps 4 and 5 were iterated using TOPCAT visualisation
interactively on smaller datasets to achive the best results.
An example is shown in Figure~\ref{fig:P3-8:allsky}.

Note, this is not necessarily the best way to prepare such maps;
executing the calculations near the data is in general more efficient
\citep{2016arXiv161109190T}.

\section{Visualisation}

\begin{figure}[ht]
\begin{center}
\includegraphics[height=0.42\textheight]{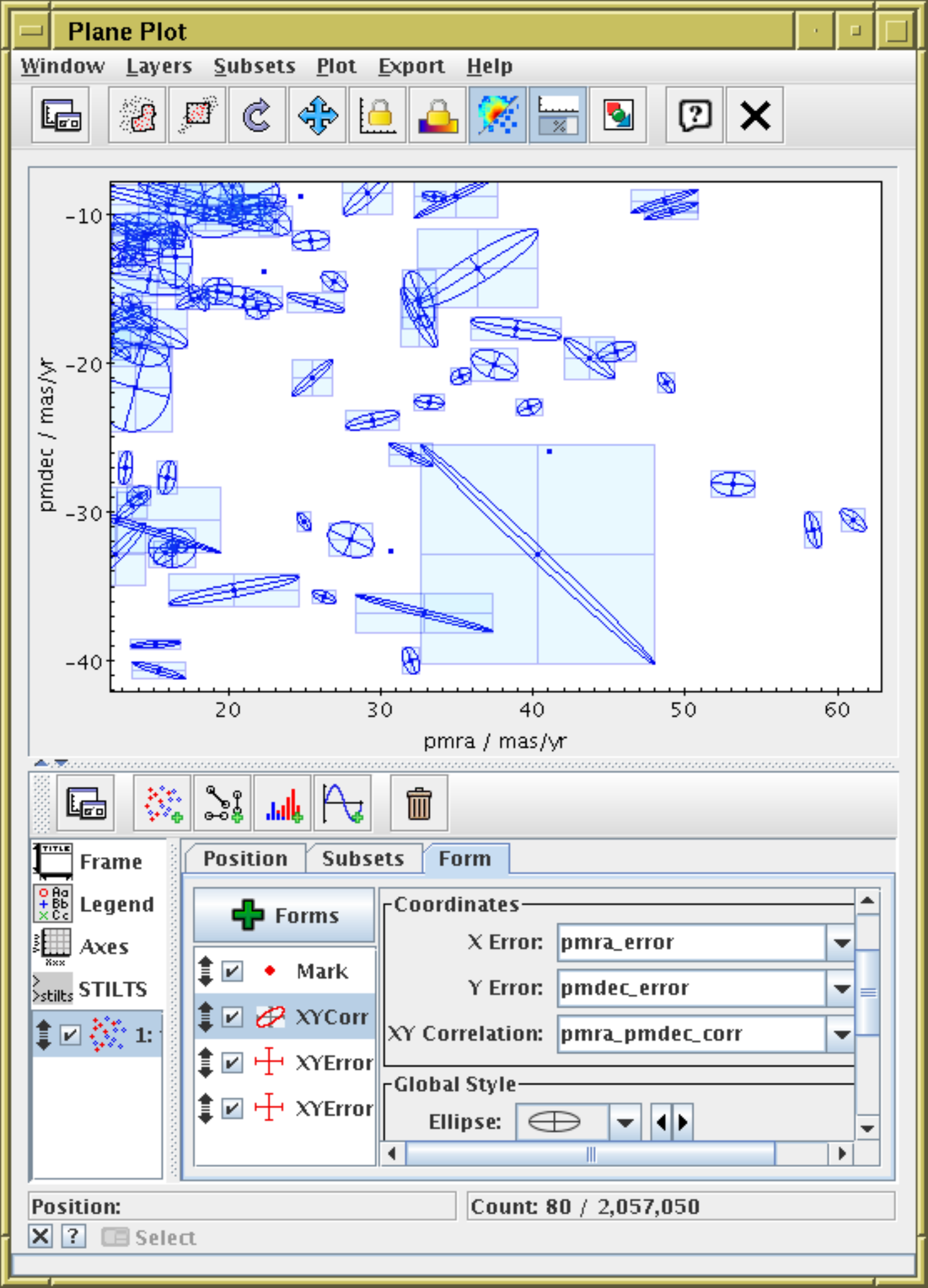}
\hspace*{0.5ex}
\includegraphics[height=0.42\textheight]{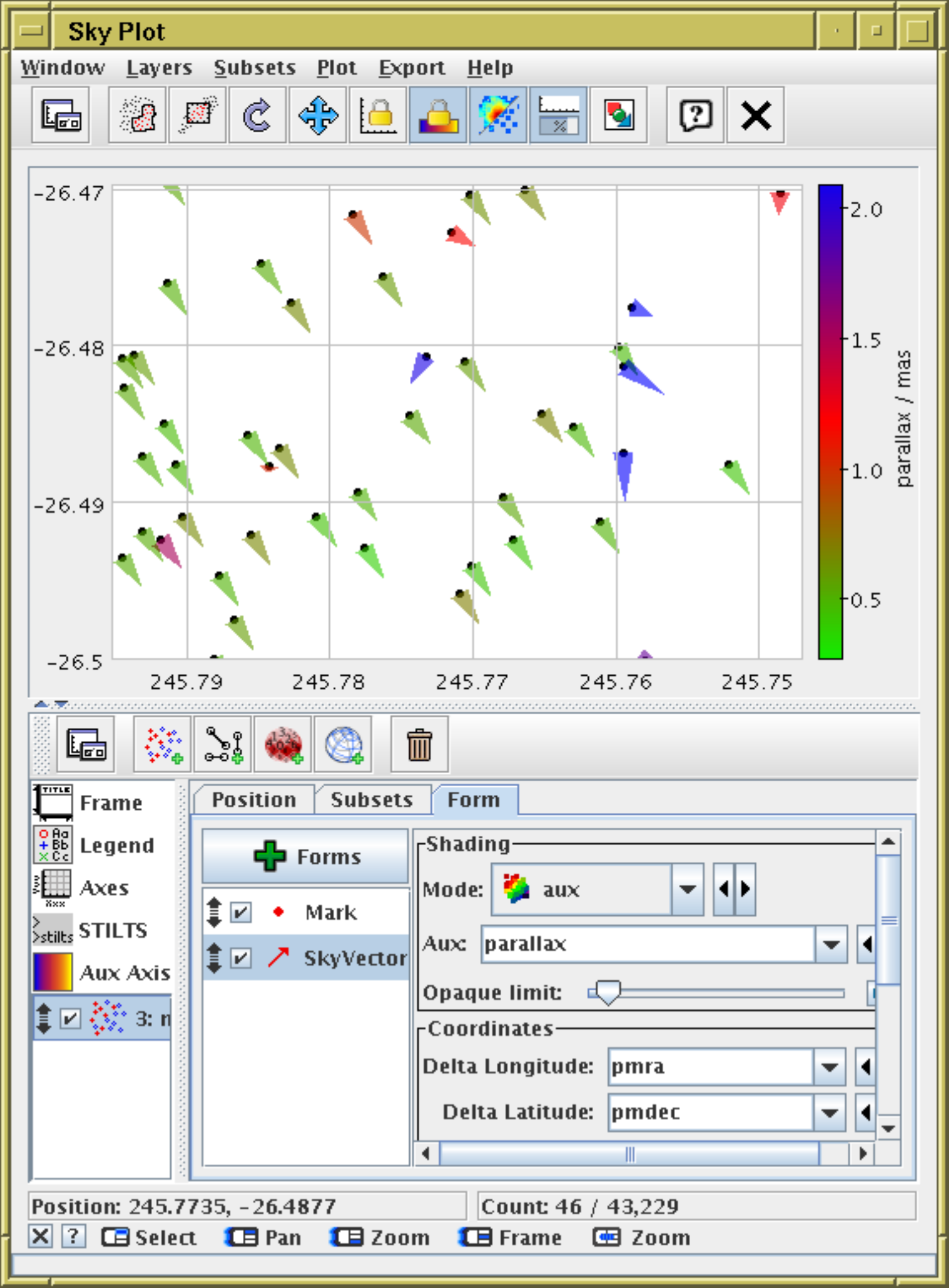}
\end{center}
\caption{Interactive visualisation of high-dimensional data in TOPCAT.
         The left hand figure displays proper motions with error ellipses
         derived from the Gaia {\tt pmra\_pmdec\_corr} column,
         which show much more information than the simple
         {\tt pmra\_error}/{\tt pmdec\_error} error boxes.
         The right hand figure shows proper montion vectors by shape,
         and parallaxes by colour.
         In each case, five dimensions are visualised.
         \label{fig:P3-8:plotshots}}
\end{figure}

TOPCAT has many visualisation modes enabling
highly configurable interactive exploration
of high-dimensional data, and suitable for
both large and small data sets.
Special attention is given to providing comprehensible representations
of large data sets --- a simple scatter plot is not useful when
there are many more points than pixels.
Two plot types have been specifically introduced or enhanced
for {\it Gaia\/} data:
the {\em Sky Vector\/} plot displays proper motion vectors on the sky,
and the {\em Sky/XY Correlation\/} plots
show error ellipses based on the astrometric error and correlation
quantities provided in the {\it Gaia\/} catalogue;
see Figure~\ref{fig:P3-8:plotshots}.
The many non-{\it Gaia}-specific visualisation options,
too numerous to describe here, are however
in most cases the core of TOPCAT's analysis capabilities for
working with {\it Gaia} and non-{\it Gaia\/} data alike.

\acknowledgements This work has been primarily funded by the UK's Science and Technology Facilities Council.  It has made use of data from the European Space Agency (ESA) mission {\it Gaia} (\url{https://www.cosmos.esa.int/gaia}), processed by the {\it Gaia} Data Processing and Analysis Consortium (DPAC, \url{https://www.cosmos.esa.int/web/gaia/dpac/consortium}).  Special thanks to the EU Horizon 2020 project ASTERICS for funding presentation of this work at ADASS 2018.

\bibliography{P3-8}  

\end{document}